\documentclass[
aps,%
12pt,%
final,%
notitlepage,%
oneside,%
onecolumn,%
nobibnotes,%
nofootinbib,%
superscriptaddress,%
noshowpacs,%
centertags]%
{revtex4}

\usepackage{bm}
\usepackage{graphics}
\usepackage{rotating}
\usepackage{amsmath}
\usepackage{amssymb}
\usepackage{epsfig}

\begin{document}

\title{Heavy tetraquarks in the hyperspherical approach\footnote{Talk presented at 
the XXVIth International Baldin Seminar on High Energy Physics Problems "Relativistic Nuclear Physics and 
Quantum Chromodynamics", JINR, Dubna, Russia, 15-20 September, 2025}}
\author{\firstname{F.~A.} \surname{Martynenko}}
\email{f.a.martynenko@gmail.com}
\affiliation{Samara University, Samara, Russia}
\author{\firstname{A.~V.} \surname{Eskin}}
\affiliation{Samara University, Samara, Russia}
\author{\firstname{A.~P.} \surname{Martynenko}}
\affiliation{Samara University, Samara, Russia}

\begin{abstract}
Within the quark model and hyperspherical method, the bound states of four heavy quarks 
and antiquarks (tetraquarks) are investigated. In hyperradial approximation, the Schr\"odinger 
equation is reduced to a one-dimensional equation after averaging over angles in hyperspace. 
This equation is solved numerically and analytically within the variational method.
The hyperfine structure of the spectrum is calculated. To increase the accuracy of the calculation, 
corrections to the energy levels from the QCD generalization of the Breit Hamiltonian 
are taken into account.
\end{abstract}

\maketitle

\section{Introduction}

Studies of bound states of heavy quarks (mesons and baryons) have been conducted within the framework 
of the theory of strong interaction - quantum chromodynamics for many decades. During this time, various 
methods for studying bound states based on quantum field principles have been formulated. At present, 
the greatest activity in studying both the hadron mass spectrum and various reactions of their formation 
and decay within the framework of non-relativistic quantum chromodynamics and various 
versions of relativistic quark models is observed.

Along with the usual quark-antiquark and three-quark hadrons, exotic hadrons built of four, five quarks 
and antiquarks (tetraquarks, pentaquarks, etc.) have also been studied in constructing the quark model. 
The calculation of observed characteristics of such multiquark states improves our understanding of properties 
of strong interaction, including the non-perturbative aspects of QCD. The first candidate to tetraquark state
was discovered by the Belle collaboration in 2003 \cite{belle}, and since then, dozens of more candidates 
for exotic hadron states have been discovered \cite{brodsky,lebed,t14,t4a}. It can be said that a qualitatively 
new period has begun in the study of bound states of quarks and gluons. In 2020, the LHCb collaboration 
discovered a new resonant state with a mass of about 6.9 GeV \cite{lhcb}. This state was later confirmed 
by other collaborations, such as ATLAS \cite{atlas} and CMS \cite{cms,cms1}, and was considered 
a candidate for a tetraquark consisting of two $c$ quarks and two $\bar c$ antiquarks. The properties 
of this state, including mechanisms of its production and decay, are currently being studied.

Since the quantum problem with the number of particles greater than or equal to three does not have an exact 
solution, various models are used to calculate the observed characteristics of multiparticle states in QCD. 
For example, the model of quark-diquark interaction in a baryon or the model of diquark-antidiquark interaction 
in a tetraquark has become widely known. Numerical results of calculations obtained within its framework 
agree well with already known experimental data and also have predictive power for states that have not yet 
been discovered. Many results obtained in the relativistic quark model \cite{t3,t3a} subsequently 
had good experimental confirmation.

The hyperspherical harmonics method is well known for describing particle bound states in quantum theory \cite{dzhibuti,Das}. This method was developed and used earlier in \cite{simonov1,simonov2,trusov1,trusov2,apm2008} 
to describe the mass spectrum of three-particle bound states of quarks - baryons. In \cite{yusimonov}, 
the mass spectrum of pentaquarks was studied within the framework of string dynamics of quarks. 
The string model of QCD was used in \cite{nefediev} to calculate the masses of the lightest scalar 
completely charmed and completely beauty tetraquarks. It should also be noted that the hyperspherical 
harmonics method was used to study four-particle states in \cite{barnea}, and in \cite{smilga} 
the hyperspherical harmonics method was used to study vector P-wave tetraquarks.

This work extends the study of heavy tetraquarks, which we carried out within the variational method 
\cite{tetra2025}. We calculate the energy levels of specific hadronic states - tetraquarks $(cc\bar c\bar c)$, 
$(bb\bar b\bar b)$, $(cc\bar b\bar b)$, using the hyperspherical approach. Note that new experimental data 
on bound states of quarks have led to the appearance of a significant number of theoretical papers 
in which the mass spectrum of tetraquarks is studied within the framework of the nonrelativistic quark model, 
the relativistic quark model, and the QCD sum rules
\cite{trichard,t4,t5,t6,t12a,t12,t13,t2,m2,tc2,tetra1,tetra2,tetra3,tetra4} (see references to other papers 
in the review articles \cite{lebed,t4,t5}). At present, there is a certain discrepancy in theoretical 
results of calculating the masses of heavy tetraquarks, which is due to both the difference in the used models 
and the choice of numerical values of main parameters of the theory.
It is useful to note here that a very closely related problem in quantum electrodynamics of bound leptonic 
states is the study of the energy levels of positronium and muonium molecules \cite{varga1,bubin}.

\section{Solution of the problem of four-particle bound state.}

Let us consider the description of the bound states of four particles in the coordinate representation. 
Let ${\bf r}_1$, ${\bf r}_2$, ${\bf r}_3$, ${\bf r}_4$ be the radius vectors of the particles in the 
initial reference frame. It is convenient to move on to the Jacobi coordinates $\boldsymbol\rho$, 
$\boldsymbol\lambda$, $\boldsymbol\sigma$, which are related to the initial coordinates as follows:
\begin{eqnarray}
{\bf r}_1=-\frac{m_2}{m_{12}}\boldsymbol\rho+\frac{m_{34}}{m_{1234}}\boldsymbol\sigma,~~~
{\bf r}_2=\frac{m_1}{m_{12}}\boldsymbol\rho+\frac{m_{34}}{m_{1234}}\boldsymbol\sigma,\\
{\bf r}_3=-\frac{m_4}{m_{34}}\boldsymbol\lambda-\frac{m_{12}}{m_{1234}}\boldsymbol\sigma,~~~
{\bf r}_4=\frac{m_3}{m_{34}}\boldsymbol\lambda-\frac{m_{12}}{m_{1234}}\boldsymbol\sigma.
\end{eqnarray}

\begin{figure}[htbp]
\centering
\includegraphics[scale=0.8]{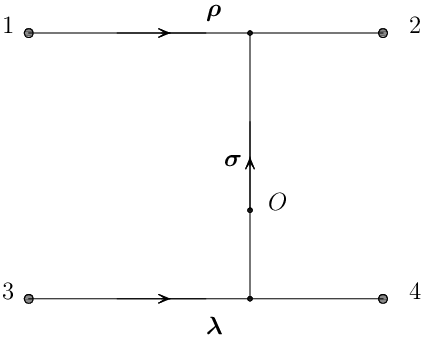}
\caption{The Jacobi coordinates in a four-particle system.}
\label{ris1}
\end{figure}

This choice of coordinates is shown in Fig.~\ref{ris1}. The coordinate $\boldsymbol\rho$ represents 
relative distance between the first pair of particles 1 and 2. The coordinate $\boldsymbol\lambda$ 
represents the relative distance between the second pair of particles 3 and 4. 
Hereafter we consider particles 1-2 are quarks, 3-4 are antiquarks.
The coordinate $\boldsymbol\sigma$ represents the relative distance between the centers of mass 
of the first and second pair.

In the Jacobi coordinates, the kinetic energy operator of the system has the form:
\begin{equation}
\label{f1}
\hat T=\frac{{\bf p}^2_\rho}{2\mu_1}+\frac{{\bf p}^2_\lambda}{2\mu_2}+\frac{{\bf p}^2_\sigma}{2\mu_3},
\end{equation}
where the reduced masses $\mu_1$, $\mu_2$, $\mu_3$ are equal to
\begin{equation}
\label{f2}
\mu_1=\frac{m_1 m_2}{m_{12}},~~~\mu_2=\frac{m_3 m_4}{m_{34}},~~~\mu_3=\frac{m_{12} m_{34}}{m_{1234}}.
\end{equation}

In nonrelativistic approximation, the total Hamiltonian of four particles includes the Coulomb 
interaction and the confinement potential. The Coulomb potential for four particles is determined 
by the sum of pairwise Coulomb interactions, and we choose the confinement potential as the sum 
of pairwise confinement potentials for the quark-quark, quark-antiquark subsystems in the form:
\begin{equation}
\label{f2a}
\Delta V^C({\bf r}_1,{\bf r}_2,{\bf r}_3,{\bf r}_4)=-\frac{4}{3}\sum_{i<j}
\frac{\alpha_{s~ij}}{|{\bf r}_i-{\bf r}_j|},
\end{equation}
\begin{equation}
\label{f3}
\Delta V^{conf}({\bf r}_1,{\bf r}_2,{\bf r}_3,{\bf r}_4)=\sum_{i<j}
A_{ij}|{\bf r}_i-{\bf r}_j|+B,
\end{equation}
where the parameters in the confinement potential are chosen in numerical calculations in exactly the same 
form as for the pair interaction of quarks and antiquarks in mesons \cite{paper1}:
$A_{ij}=0.18~GeV^2$, $\alpha_{s~c\bar c}=0.314$, $\alpha_{s~b\bar b}=0.207$, $\alpha_{s~c\bar b}=0.265$,
$m_c=1.55~GeV$, $m_b=4.88~GeV$. The strong interaction constant $\alpha_{s~ij}$ 
for a pair of quarks is half as much as the strong interaction constant for a 
quark-antiquark pair. The same applies to the confinement constant $A_{ij}$. The full value 
of the spectrum shift constant is chosen equal to $B=-0.8~GeV$.

The method of hyperspherical functions is widely used in the study of multiparticle bound states. 
Let us consider it for the case of four particles, limiting ourselves to studying  the ground states 
of tetraquarks without radial or orbital excitations. Let us pass to the hyperspace of dimension 9, 
which consists of the Jacobi variables $\boldsymbol\rho$, $\boldsymbol\lambda$, $\boldsymbol\sigma$. 
Denoting the hyperradius by $R$, we represent the relationship of the radial variables $\rho$, $\lambda$, 
$\sigma$ with $R$ as follows:
\begin{equation}
\label{f4}
\rho=R\sin\theta\cos\phi,~~~\lambda=R\sin\theta\sin\phi,~~~\sigma=R\cos\theta.
\end{equation}

Then the volume element in hyperspace of dimension nine has the form:
\begin{equation}
\label{f5}
dV_9=(4\pi)^3R^8\sin^2\phi\cos^2\phi\sin^5\theta\cos^2\theta dRd\theta d\phi.
\end{equation}

The volume of an 8-dimensional sphere is:
\begin{equation}
\label{f6}
\Omega_9=(4\pi)^3\int_0^{\pi/2}\sin^2\phi\cos^2\phi d\phi
\int_0^{\pi/2}\sin^5\theta\cos^2\theta d\theta =\frac{32\pi^4}{105}.
\end{equation}

When moving into hyperspace, it is convenient to transform the kinetic energy operator by performing 
an additional scaling transformation of the Jacobi variables:
\begin{equation}
\label{f7}
\boldsymbol\rho_{ij}\to \sqrt{\frac{\mu}{\mu_1}}\boldsymbol\rho_{ij},~~~
\boldsymbol\lambda_{ij}\to \sqrt{\frac{\mu}{\mu_2}}\boldsymbol\lambda_{ij},~~~
\boldsymbol\sigma_{ij}\to \sqrt{\frac{\mu}{\mu_3}}\boldsymbol\sigma_{ij},
\end{equation}
where $\mu$ is an auxiliary mass parameter, the dependence on which disappears in the final expression 
for the binding energy. The indices of the Jacobi variables represent the indices of the particles and 
are written here in the same way.

As a result, the kinetic energy operator in the center of mass system will take the form:
\begin{equation}
\label{f8}
\hat T=-\frac{1}{2\mu}\left(\frac{\partial^2}{\partial R^2}+\frac{8}{R}\frac{\partial}
{\partial R}+
\frac{K^2(\Omega)}{R^2}\right),
\end{equation}
where $K^2(\Omega)$ is the square of the 9-dimensional angular momentum. The hyperspherical 
functions are eigenfunctions for $K^2(\Omega)$:
\begin{equation}
\label{f9}
K^2(\Omega)Y_{[K]}=K(K+7)Y_{[K]}.
\end{equation}

Any coordinate wave function can be expanded in the complete set of hyperspherical 
functions.
The transformations performed during the transition to hyperspace and the resulting formula \eqref{f8} 
ultimately lead to a one-dimensional radial Schr\"odinger equation under simplifying assumptions 
about the form of the tetraquark wave function.

The Jacobi coordinates completely determine the positions of the four particles. By changing the numbering 
of the particles, we obtain different sets of coordinates ($\boldsymbol\rho_{12}$, $\boldsymbol\lambda_{34}$, 
$\boldsymbol\sigma_{12,34}$, etc.). Omitting the particle indices in the Jacobi variables in what follows, 
we represent the square of the hyperradius $R^2$ by the following sum:
\begin{equation}
\label{f10}
R^2=\boldsymbol\rho^2+ \boldsymbol\lambda^2+\boldsymbol\sigma^2.
\end{equation}

Since we are now considering the ground states of the tetraquarks, the eigenvalue of the square of the angular 
momentum $K$ is assumed to be zero. The general solution of the original Schr\"odinger equation can be represented 
as an expansion in hyperspherical functions in the symbolic form:
\begin{equation}
\label{f11}
\psi(\boldsymbol\rho,\boldsymbol\lambda,\boldsymbol\sigma)=\sum_{K}\psi_K(R)Y_{[K]}(\Omega).
\end{equation}

In real calculations, it is necessary to find the hyperradial functions numerically. To do this, 
\eqref{f11} is substituted into the original Schr\"odinger equation, which yields a set of non-homogeneous 
hyperradial differential equations.

\section{Hyperradial approximation.}

In the hyperradial approximation $K=0$ and the wave function of the particle system does not depend 
on the angular variables: $\psi=\psi(R)$. This circumstance allows us to perform
averaging over the angular variables in the original Schr\"odinger equation.

Averaging over the angles in the Coulomb and confinement terms of the potential takes the form:
\begin{equation}
\label{f12}
<\frac{1}{\rho_{ij}}>=\frac{(4\pi)^3}{R\Omega_9}\int_0^{\pi/2}d\theta\int_0^{\pi/2}d\phi \sin^2\phi \cos\phi \sin^4\theta
\cos^2\theta =\frac{35}{16}\frac{1}{R}=<\frac{1}{\lambda_{ij}}>,
\end{equation}
\begin{equation}
\label{f13}
<\rho_{ij}>=\frac{R(4\pi)^3}{\Omega_9}\int_0^{\pi/2}d\theta\int_0^{\pi/2}d\phi \sin^2\phi 
\cos^3\phi \sin^6\theta\cos^2\theta =\frac{35}{64}R=<\lambda_{ij}>.
\end{equation}

By introducing the reduced wave function $\chi(R)=R^4\psi(R)$ and averaging the interaction potential 
of quarks and antiquarks over an eight-dimensional sphere, we obtain the following Schr\"odinger equation:
\begin{equation}
\label{f14}
\frac{d^2\chi(R)}{dR^2}+2\mu\left[E+\frac{a}{R}-bR-\frac{6}{\mu R^2}\right]\chi(R)=0.
\end{equation}
The wave function $\chi(R)$ depends only on the hyperradius and is symmetric under the permutation 
of quarks and antiquarks in the tetraquark. Note that the centrifugal potential in equation \eqref{f14} 
is not equal to zero even in the case of zero moment $K$.

In the Schr\"odinger equation for the reduced wave function $\chi(R)$ there is an averaged over angles 
potential of interaction of quarks and antiquarks:
\begin{equation}
\label{f15}
V(R)=<V(\boldsymbol\rho,\boldsymbol\lambda,\boldsymbol\sigma)>=\int(V^C+V^{conf})
\frac{d\Omega}{\Omega_9}.
\end{equation}

By further introducing a new variable $x=R\sqrt{\mu}$, we can transform the Schr\"odinger equation 
\eqref{f14}, eliminating from it the dependence on the mass $\mu$:
\begin{equation}
\label{f16}
\frac{d^2\chi(x)}{dx^2}+2\left[E+\frac{a}{x}-bx-\frac{6}{x^2}\right]\chi(x)=0.
\end{equation}

The effective constants $a$, $b$ included in the Coulomb and confinement parts of the potential in \eqref{f16} 
do not depend on $\mu$ and are determined by the following expressions for tetraquarks consisting of identical 
quarks and antiquarks with mass $m$:
\begin{equation}
\label{f17}
a=\frac{175}{12\sqrt{2}}\alpha_s\sqrt{m},~~~b=\frac{175}{32\sqrt{2}}\frac{A}{\sqrt{m}}.
\end{equation}

The resulting one-dimensional Schr\"odinger equation for $\chi(R)$ is solved in two ways: numerically 
using the Mathematica program \cite{lucha} and analytically using the variational method. 
To solve the problem using the variational method, we introduce a trial function of the form:
\begin{equation}
\label{f18}
\chi(x)=\frac{2}{\sqrt{5}}p^9 x^4 e^{-p^3 x^{3/2}}.
\end{equation}
Usually, either a basis of hydrogen functions or a Gaussian basis is used to construct variational functions. 
The dependence on the $x$ coordinate in the exponential \eqref{f18} is somewhat modified, since 
in the asymptotics for $x\to\infty$ the equation \eqref{f16} has the form of the well-known Airy equation:
\begin{equation}
\label{f19}
\frac{d^2\chi(x)}{dx^2}-2bx\chi(x)=0.
\end{equation}
The behavior at infinity of the solution of this equation is determined by the formula:
\begin{equation}
\label{f20}
Ai(z)\sim \frac{1}{2\sqrt{\pi}}z^{-1/4}e^{-\frac{2}{3}z^{3/2}},~~~z=(2b)^{1/3}x.
\end{equation}
which allows us to further use a variational function of the form \eqref{f18}. Further calculations 
show the preference of such a choice of variational function in comparison with other options (hydrogen 
or Gaussian wave function).

The dependence of the binding energy on variational parameter $p$ is then determined by the following 
expression:
\begin{equation}
\label{f21}
E=\frac{119\Gamma(\frac{14}{3})}{480\cdot 2^{2/3}}p^4-\frac{\Gamma(\frac{16}{3})}{60\cdot 2^{1/3}}ap^2+
\frac{\Gamma(\frac{20}{3})}{120\cdot 2^{2/3}}\frac{b}{p^2}.
\end{equation}

The variational parameter itself is found from the solution of the equation:
\begin{equation}
\label{f22}
119 \Gamma(\frac{14}{3})p^3-2b\Gamma(\frac{20}{3})\frac{1}{p^3}-4a\cdot 2^{1/3}\Gamma(\frac{16}{3})p=0.
\end{equation}

Since analytical solution of this equation looks rather cumbersome, we will immediately present here its 
numerical solution, which is obtained for the tetraquark $(cc\bar c\bar c)$: $p_0=0.909983391$ GeV$^{1/4}$. 
Corresponding binding energy for this tetraquark is $E_0=0.382$ GeV. It is useful to compare this value 
with the binding energy $E_0=0.347$ GeV obtained in \cite{tetra2025} within the framework of the variational 
method and with numerical value of the binding energy $E_0=0.380$ GeV obtained using the Mathematica 
program \cite{lucha}. Despite the simplicity of variational solution of the problem, good agreement 
is obtained with more accurate numerical values, which agree well with each other.

\begin{figure}[htbp]
\centering
\includegraphics[scale=0.8]{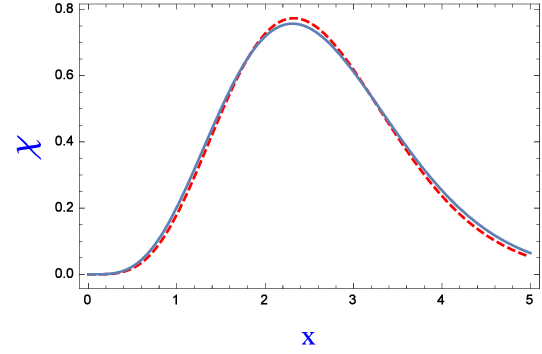}
\includegraphics[scale=0.8]{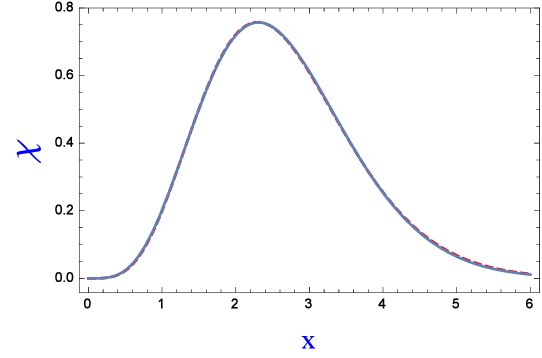}
\caption{Tetraquark wave function in the hyperradial approximation. The solid line represents 
numerical solution of the Schr\"odinger equation, and the dotted line corresponds to the variational 
wave function \eqref{f18}. The left figure shows the function \eqref{f18}, and the right figure shows 
the function \eqref{f23}.}
\label{ris2}
\end{figure}

It is also useful to compare the form of variational wave function \eqref{f18} and numerical solution 
of the problem (for the wave function) in the Mathematica package \cite{lucha}, which is presented 
in left Fig.~\ref{ris2}. It follows from the graph that these two wave functions almost coincide, 
which is somewhat unexpected, since only one variational parameter is used in \eqref{f18}. In the case 
of the hydrogen wave function, or the Gaussian wave function, which have a different asymptotics 
at infinity, there is a difference from numerical solution of the Schr\"odinger equation, which 
manifests itself both in the region of small and large $x$. In this regard, it can be stated that 
the choice of the asymptotic behavior of $\chi(x)$ in \eqref{f18} turned out to be very successful. 
To further clarify the results obtained, two variational parameters $p$ and $q$ can be introduced, 
representing the wave function in the form:
\begin{equation}
\label{f23}
\chi(x)=\sqrt{\frac{2^{\frac{9}{q}} q p^{\frac{9}{q}}}{\Gamma(\frac{9}{q})}}
x^4 e^{-p x^{q}}.
\end{equation}

\begin{figure}[htbp]
\centering
\includegraphics[scale=0.8]{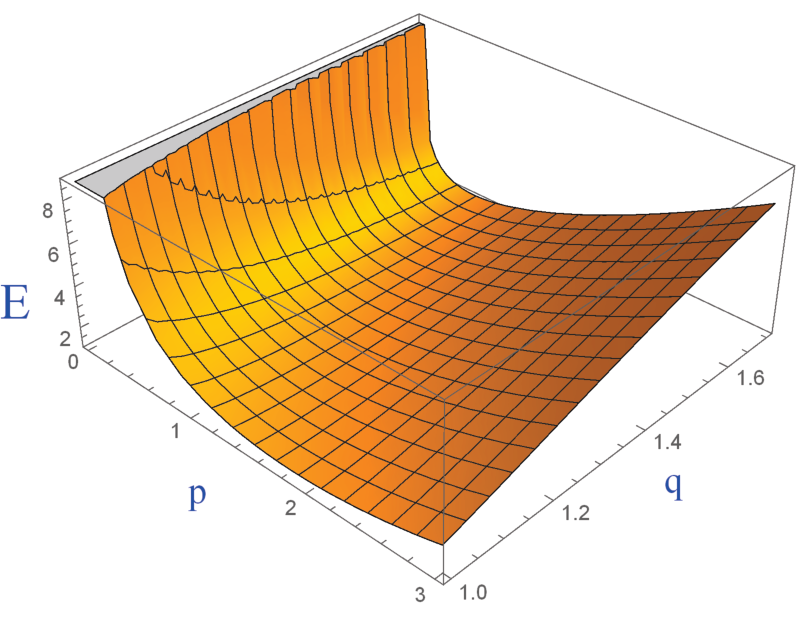}
\caption{Binding energy as a function of two variational parameters $p$, $q$.}
\label{ris}
\end{figure}

The dependence of the binding energy on these two parameters is shown in Fig.~\ref{ris}. Minimization 
of the Hamiltonian yields the following values of the parameters and the binding energy:
$p=0.9467404357$ GeV$^{\frac{q}{2}}$, $q=1.3610537691$, $E=0.380$ GeV. The binding energy value 
has not changed compared to \eqref{f18} with an accuracy of three digits after the decimal point. 
In this case, variational wave function \eqref{f23} practically coincides with numerical 
wave function, as shown in the right Fig.~\ref{ris2}. The resulting analytical expression for the tetraquark 
wave function \eqref{f23} allows it to be used to calculate a number of corrections in the tetraquark energy 
spectrum using perturbation theory. Note also that the value of obtained tetraquark wave function $\Psi_T(0)$ 
using \eqref{f23} is equal to $\Psi_T(0)=0.09~GeV^{9/2}$, which is in good agreement with the value 
$0.10~GeV^{9/2}$ obtained within the framework of a more rigorous variational method and used 
to estimate the probabilities of tetraquark production in the Higgs boson decay in \cite{tt2025,tet2025}.

\section{Hyperfine structure of the tetraquark spectrum.}

The hyperfine structure of the tetraquark spectrum is determined by the pair potentials of the spin-spin, 
spin-orbit interaction of the quarks and antiquarks that form the tetraquark. In the case of the ground state, 
the contribution to the hyperfine splitting is given by the following interaction operator, which is determined 
by the one-gluon exchange:
\begin{equation}
\label{f24}
\Delta V^{hfs}({\boldsymbol\rho},{\boldsymbol\lambda},{\boldsymbol\sigma})=
a_{12}({\bf s}_1{\bf s}_2)+a_{34}({\bf s}_3{\bf s}_4)+a_{13}({\bf s}_1{\bf s}_3)+
a_{14}({\bf s}_1{\bf s}_4)+a_{23}({\bf s}_2{\bf s}_3)+a_{24}({\bf s}_2{\bf s}_4),
\end{equation}
\begin{eqnarray}
\label{f25}
a_{12}=\frac{16\pi\alpha_{s12}}{9m_1m_2}\delta({\bf r}_{12}),~~~
a_{34}=\frac{16\pi\alpha_{s34}}{9m_3m_4}\delta({\bf r}_{34}),~~~
a_{13}=\frac{32\pi\alpha_{s13}}{9m_1m_3}\delta({\bf r}_{13}),\\
a_{14}=\frac{32\pi\alpha_{s14}}{9m_1m_4}\delta({\bf r}_{14}),~~~
a_{23}=\frac{32\pi\alpha_{s23}}{9m_2m_3}\delta({\bf r}_{23}),~~~
a_{24}=\frac{32\pi\alpha_{s24}}{9m_2m_4}\delta({\bf r}_{24}),
\end{eqnarray}
where ${\bf s}_i$ $(i=1, 2, 3, 4)$ are the spin operators of quarks and antiquarks, $\alpha_{s~ij}$ 
is the constant of strong interaction of particles $i$ and $j$.

The spin wave function of the tetraquark $\chi_{SS_z}^{S_{12}S_{34}}$ is symmetric with respect 
to the permutation of two quarks 1 and 2 and two antiquarks 3 and 4 ($S_{12}=1$ is the spin 
of the quark pair, $S_{34}=1$ is the spin of the antiquark pair):
\begin{eqnarray}
\label{f26}
\chi_{00}^{11}=\frac{1}{\sqrt{12}}\left(2\uparrow\uparrow\downarrow\downarrow-
\uparrow \downarrow   \uparrow \downarrow - \uparrow  \downarrow\downarrow \uparrow-
\downarrow\uparrow\uparrow\downarrow-\downarrow \uparrow \downarrow\uparrow+
2 \downarrow\downarrow \uparrow\uparrow\right),  \\
\chi_{11}^{11}=\frac{1}{2}\left(\uparrow\uparrow\uparrow\downarrow+ \uparrow\uparrow  \downarrow\uparrow-
\uparrow \downarrow \uparrow\uparrow-\downarrow\uparrow\uparrow\uparrow\right),  \\
\chi_{22}^{11}=\uparrow\uparrow\uparrow\uparrow,~~~\uparrow={1\choose{0}},~~~\downarrow={0\choose{1}}.
\end{eqnarray}

The permutation of identical quarks 1 and 2, as well as the permutation of two identical antiquarks 3 and 4, 
must change the sign of the total wave function. Therefore, the color part of the tetraquark wave function 
is antisymmetric when two quarks or two antiquarks are permuted.

To calculate matrix elements with $\delta$-functions, we use the explicit form of the tetraquark wave 
function \eqref{f23}. The first matrix element $<\delta({\bf r}_1-{\bf r}_2)>$ has the form:
\begin{equation}
\label{f27}
<\delta({\bf r}_{12})>=<\delta({\boldsymbol \rho})>=\frac{105}{\pi}
\frac{2^{(\frac{3}{q}-\frac{13}{2})}p^{\frac{3}{q}}\Gamma(\frac{6}{q})m^{\frac{3}{2}}}
{\Gamma(\frac{9}{q})}=\delta=0.0846257426~GeV^{3},
\end{equation}
where the numerical value of this matrix element for the tetraquark $(cc\bar c\bar c)$ is given. When calculating 
the coordinate matrix elements with other delta functions $<\delta({\bf r}_{13})>$,
$<\delta({\bf r}_{14})>$, $<\delta({\bf r}_{23})>$, $<\delta({\bf r}_{24})>$ we can use the above-discussed 
permutation of particle coordinates when choosing the basic variables $\boldsymbol\rho$, 
$\boldsymbol\lambda$, 
$\boldsymbol\sigma$ and the symmetry of the wave function \eqref{f23}. Thus, it turns out that all matrix
elements with $\delta$-functions coincide and are equal to $\delta$ \eqref{f27}. Then the contribution 
to the hyperfine structure of the tetraquark spectrum is determined by the expression:
\begin{equation}
\label{f28}
\Delta E^{hfs}=\kappa<({\bf s}_1{\bf s}_2)>+\kappa<({\bf s}_3{\bf s}_4)>+
2\kappa<({\bf s}_1{\bf s}_3)>+2\kappa<({\bf s}_1{\bf s}_4)>+
\end{equation}
\begin{displaymath}
2\kappa<({\bf s}_2{\bf s}_3)>+
2\kappa<({\bf s}_2{\bf s}_4)>=\frac{\kappa}{2}\bigl[2S_T(S_T+1)-7\bigr],~~~
\kappa=\frac{35\alpha_s}{3\sqrt{m}}\frac{2^{\frac{3}{q}-\frac{5}{2}}p^{\frac{3}{q}}\Gamma(\frac{6}{q})}
{\Gamma(\frac{9}{q})}
\end{displaymath}
where the spins of the pair of quarks $s_{12}$ and antiquarks $s_{34}$ are chosen to be equal to 1.

Since the spin of a tetraquark takes three values 0, 1, 2, the corresponding energy values are:
\begin{eqnarray}
\label{f29}
E^{hfs}(S_T=0)=E_0-\frac{7}{2}\kappa,~~~E^{hfs}(S_T=1)=E_0-\frac{3}{2}\kappa,~~~
E^{hfs}(S_T=2)=E_0+\frac{5}{2}\kappa.
\end{eqnarray}
The hyperfine splitting in \eqref{f29} is determined by the quantity $\kappa$.

When adding up the spins of quarks, we use the following scheme: first, we add up the spins of the first 
and second $c$-quarks, the third and fourth $\bar c$-antiquarks. In this case, the total spins of these pairs 
are $S_{cc}=1$, $S_{\bar c\bar c}=1$ since the quarks are identical, and the antisymmetry of the wave function 
of the system is achieved due to its color part. Then we add up the two spins ${\bf S}_{cc}+{\bf S}_{\bar c\bar c}$ 
and we obtain that the total angular momentum of the tetraquark in the ground state takes the values 
0, 1, 2.

The charge and spatial parities of a tetraquark are:
\begin{eqnarray}
\label{f30}
C_T=(-1)^{S_T+L_T},~~~P_T=(-1)^{L_T}.
\end{eqnarray}

The S-wave states of a tetraquark with such a sequence of addition of quark moments are determined 
by the expressions:
\begin{eqnarray}
\label{hfs9}
|0^{++}>_{T}=|S_{cc}=1,S_{\bar c\bar c}=1,S_T=0,L_T=0>_{J_T=0},\\
|1^{+-}>_{T}=|S_{cc}=1,S_{\bar c\bar c}=1,S_T=1,L_T=0>_{J_T=1},\\
|2^{++}>_{T}=|S_{cc}=1,S_{\bar c\bar c}=1,S_T=2,L_T=0>_{J_T=2}.
\end{eqnarray}

\begin{table}[htbp]
\caption{Ground state masses of tetraquarks in GeV.}
\bigskip
\label{tb1}
\begin{tabular}{|c|c|c|c|c|c|c|c|c|c|} \hline
State&$(cc\bar c\bar c)$&\cite{tetra2025}&\cite{t3}&\cite{t2}&\cite{t12a} &\cite{t12}&\cite{m2}& \cite{t13} &\cite{tc2} \\ \hline
$0^{++}$&5.86 &6.10 &6.190&6.477&5.966 & 5.9694&5.883& 6.435  &  6.503    \\  \hline
$1^{+-}$&6.02 &6.26 &6.271&6.528&6.051 &6.0209&6.120&  6.515  & 6.517       \\  \hline
$2^{++}$&6.35 &6.57 &6.367&6.573& 6.223&6.1154& 6.246&  6.543 &  6.544      \\  \hline
\end{tabular}
\begin{tabular}{|c|c|c|c|c|c|c|c|c|c|} \hline
State&$(bb\bar b\bar b)$ &\cite{tetra2025}&\cite{t3}& \cite{t12a}&\cite{m2} & \cite{t13} \\ \hline
$0^{++}$& 18.63&18.81 &19.314 &18.754  &18.748 & 19.201   \\  \hline
$1^{+-}$&18.76 &18.86&19.320  &18.808 &18.828 &   19.251    \\  \hline
$2^{++}$& 19.02&18.97 &19.330 &18.916  &18.900 &  19.262    \\  \hline
\end{tabular}
\begin{tabular}{|c|c|c|c|c|c|} \hline
State&$(cc\bar b\bar b)$ &\cite{tetra2025} &\cite{t3} & \cite{m2}&  \cite{t13} \\ \hline
$0^{++}$&12.44 &12.77&12.846 & 12.445 &  13.496  \\  \hline
$1^{+-}$&12.47 &12.78 & 12.859 & 12.536 & 13.560   \\  \hline
$2^{++}$&12.52 &12.80&12.883 & 12.614 & 13.595  \\  \hline
\end{tabular}
\end{table}

To improve the accuracy of the calculation, we also calculated a number of basic corrections 
in the energy spectrum, which are determined by the QCD generalization of the Breit potential.

\section{Corrections in the tetraquark energy spectrum}

The correction for relativity is one of main corrections in the energy spectrum. This applies in particular 
to the motion of $c$-quarks. General expression for this correction in the Hamiltonian of the system 
is equal to the sum of terms for each quark and antiquark of the following form:
\begin{equation}
\label{f31}
\hat T_{rel}=-\frac{{\bf p}^4_1}{8m_1^3}-\frac{{\bf p}^4_2}{8m_2^3}-\frac{{\bf p}^4_3}{8m_3^3}
-\frac{{\bf p}^4_4}{8m_4^3}.
\end{equation}

When calculating the corresponding matrix elements for quarks and antiquarks of the same mass, 
we express the momenta ${\bf p}_i$ through derivatives with respect to the Jacobi coordinates:
\begin{equation}
\label{f32}
\nabla_{{\bf r_1}}=\nabla_{\boldsymbol\rho}-\frac{1}{2}\nabla_{\boldsymbol\sigma},~
\nabla_{{\bf r_2}}=\nabla_{\boldsymbol\rho}+\frac{1}{2}\nabla_{\boldsymbol\sigma},~
\nabla_{{\bf r_3}}=\nabla_{\boldsymbol\lambda}+\frac{1}{2}\nabla_{\boldsymbol\sigma},~
\nabla_{{\bf r_4}}=\nabla_{\boldsymbol\lambda}-\frac{1}{2}\nabla_{\boldsymbol\sigma}.
\end{equation}

Calculating analytically the matrix element of the operator \eqref{f31} by wave functions $\psi(R)$ 
from \eqref{f23}, we obtain the following result:
\begin{equation}
\label{f33}
\Delta E_{rel}=-\frac{5\cdot 2^{(\frac{4}{q}-9)} (90q^3+675 q^2+1250 q+20337)p^{\frac{4}{q}}\Gamma(\frac{5}{q})}
{33m\Gamma(\frac{9}{q})}.
\end{equation}

The second correction in the Hamiltonian of the system, similar in structure, is the relativistic recoil correction. 
We will represent corresponding operator in the Hamiltonian, expressing it in terms of relative momenta 
of particles ${\bf p}_{ij}$:
\begin{equation}
\label{f34}
\delta H_{rel-rec}=-\sum_{i,j}c_{ij}\frac{\alpha_{s~ij}}{3m_i m_j r_{ij}}
\left[{\bf p}^2_{ij}+\frac{{\bf r}_{ij}({\bf r}_{ij}{\bf p}_{ij}){\bf p}_{ij}}{r^2_{ij}}\right],
\end{equation}
where the coefficients $c_{ij}$ are equal to 1 for a pair of quarks or a pair of antiquarks $i$, $j$ and 2 
in the case of a quark-antiquark pair. Calculating matrix element of the operator \eqref{f34}, which 
is also performed analytically, yields the result:
\begin{equation}
\label{f35}
\Delta E_{rel-rec}=-\frac{175\alpha_s 2^{(\frac{3}{q}-\frac{15}{2})} q^2 p^{\frac{3}{q}}\Gamma(2+\frac{6}{q})}
{3\sqrt{m}\Gamma(\frac{9}{q})}.
\end{equation}
In obtaining it, we used, as before, the symmetry of the tetraquark wave function when permuting 
the particle coordinates and the possibility of different choices of the Jacobi coordinates. The corrections 
\eqref{f33} and \eqref{f35} give a significant negative shift in the energy spectrum.

We also take into account the correction for contact interaction from the Breit potential, which is 
determined in the Hamiltonian by the sum of pairwise interactions according to the formula:
\begin{equation}
\label{f36}
\delta H_{cont}=\sum_{i,j}c_{ij}\frac{\pi\alpha_{s~ij}}{3}\left(\frac{1}{m_i^2}+\frac{1}{m_j^2}\right)
\delta({\bf r}_{ij}).
\end{equation}

In this case, the matrix element is determined by the same $\delta$-function as in the case of hyperfine 
interaction and is expressed through variational parameters $p$, $q$ in the form:
\begin{equation}
\label{f37}
\Delta E_{cont}=\frac{175\alpha_s 2^{(\frac{3}{q}-\frac{9}{2})} p^{\frac{3}{q}}\Gamma(\frac{6}{q})}
{\sqrt{m}\Gamma(\frac{9}{q})}.
\end{equation}

To clarify the value of hyperfine splitting in quarkonia, a nonperturbative potential with spin-spin 
interaction of the form \cite{repko1,repko2} is used:
\begin{equation}
\label{f38}
\Delta H^{hfs}_{conf}=\sum_{i,j}
f_V\frac{A}{8r_{ij}}\left\{\frac{1}{m_i^2}+\frac{1}{m_j^2}+\frac{16}{3m_im_j}({\bf s}_i{\bf s}_j)\right\},
\end{equation}
where another parameter $f_V=0.9$ is introduced. Let us separately represent the contribution of this 
operator for states with spin $S_T=0, 1, 2$:
\begin{equation}
\label{f39}
\Delta E^{hfs}_{conf}=
\begin{cases}
S_T=0, \frac{A f_V}{(2m)^{3/2}}\lambda,~~~\lambda=\frac{35 \cdot 2^{(\frac{1}{q}-7)} q p^{\frac{1}{q}}
\Gamma(\frac{q+8}{q})}
{\Gamma(\frac{9}{q})}\\
S_T=1, \frac{A f_V}{(2m)^{3/2}} \frac{7}{3}\lambda\\
S_T=2,  \frac{A f_V}{(2m)^{3/2}} 5\lambda.
\end{cases}
\end{equation}

Thus, total values of the tetraquark $(cc\bar c\bar c)$ $0^{++}$, $1^{+-}$, $2^{++}$ masses are determined 
by the sum of expressions \eqref{f29}, \eqref{f33}, \eqref{f35}, \eqref{f37}, \eqref{f39}
which are functions of two variational parameters $p$, $q$.
The results of numerical calculation of the mass spectrum of heavy tetraquarks are presented in Table~\ref{tb1}.
It also shows some calculation results from other works.

\section{Conclusion}

Recently, a number of candidates for the role of non-standard multiquark mesons and baryons containing heavy 
$c$ and $b$ quarks have been discovered. The number of experimental works on exotic hadrons is steadily growing, 
their composition and properties are being studied. In the analysis of these experimental data, various 
theoretical estimates of the masses of such states obtained within the framework of QCD models play 
an important role. The stage of identifying exotic hadrons and studying the mechanisms that lead 
to their formation is coming. So far, it cannot be said that we understand the nature of the newly 
discovered resonance states well, which casts doubt on the understanding of hadron structures in general.

The quark model used in this work for calculating the mass spectrum of heavy tetraquarks is based 
on the hyperspherical expansion method. For the ground states of tetraquarks, the hyperradial approximation 
is used, in which the wave function is independent of angular variables in hyperspace. This model allows 
one to obtain a simple expression for the tetraquark wave function, which depends on the hyperradius, 
in an analytical form in the nonrelativistic approximation. Based on this analytical solution, the results 
for the energy levels are refined by calculating the main corrections, which are determined by the 
generalization of the Breit potential in quantum chromodynamics. Thus, when calculating the masses 
of the ground states tetraquarks, we take a Hamiltonian consisting of terms known from previous 
calculations of meson and baryon masses. All used parameters of the quark model are chosen 
to be the same in magnitude as in previous calculations, which, as is known, gave values 
of observed quantities with an accuracy of 0.1 percent.

The tetraquark $T_{cc\bar c\bar c}$ includes two pairs of identical particles. In its construction, 
we consider these pairs in an antisymmetric color state, which, with a symmetric coordinate wave function 
and spin wave function, ensures the antisymmetry of the tetraquark wave function when the identical 
particles are permuted. With such a choice of color state, an attractive potential acts between 
identical particles.

The results from different papers presented in Table~\ref{tb1} show that there is a discrepancy 
in predictions for the tetraquark masses in different models. These differences in the results 
are quite explainable and are determined primarily by different values of the quark masses in the tetraquark, 
the values of strong interaction constant, and the choice of constants in the confinement potential. 
But despite these differences, all the results obtained together should contribute to the experimental 
search for tetraquarks by the Belle, LHCb, CMS, and ATLAS collaborations. The results obtained in this work 
(see Table~\ref{tb1}) differ slightly from our results from \cite{tetra2025}, obtained within the framework 
of variational method, since the relativistic corrections and corrections with recoil, taken into 
account in this calculation, are numerically large and enter total result with a negative sign.

Our previous study of the mass spectrum of heavy tetraquarks in \cite{tetra2025} was performed within 
variational method in non-relativistic approximation. In this paper, we use a different approach 
to solving this problem, but we retained the values of all parameters of the quark model. If we compare 
the calculations of the binding energies in non-relativistic approximation in these two approaches, 
we should note the difference of 0.03 GeV (0.380 GeV and 0.347 GeV), which may mean that real tetraquark 
wave function has a small dependence on angles, which ultimately leads to this difference. Another error 
in our calculations is connected with taking into account corrections of higher order than those considered 
in Section 5. First of all, these are radiative corrections of order $O(\alpha_s)$ and 
relativistic corrections of higher order, which can be about 30 percent of those that are stated. 
Taking into account therefore the magnitude of individual effects $\sim 0.3\div 0.5$~GeV for the tetraquark 
$(cc\bar c\bar c)$ in this work, one can estimate the possible magnitude of theoretical 
calculation error at 0.1 GeV, which is about $1\div 2$ percent of the obtained mass values.

\begin{acknowledgments}
The work of F.A.M. is supported by the Foundation for the Development of Theoretical Physics 
and Mathematics BASIS (grant No. 25-1-4-15-1).
\end{acknowledgments}

\end{document}